\documentclass[amsmath,amsfonts,amssymb,twocolumn,superscriptaddress,showpacs]{revtex4-1}
\usepackage{graphicx,psfrag,color}
\usepackage{dcolumn}
\usepackage{bm}
\usepackage{mathbbol, appendix}
\usepackage{multirow}
\usepackage{booktabs}
\graphicspath {{./figures/}}
\makeatletter
\def\input@path{{./figures/}}
\makeatother

\begin{document}

\title{The Binomial Spin Glass}
\author{Mohammad-Sadegh Vaezi}
\affiliation{Department of Physics, Washington University, St. Louis, MO 63160, USA}
\author{Gerardo Ortiz}
\affiliation{Department of Physics, Indiana University, Bloomington,
IN 47405, USA}
\affiliation{Department of Physics, University of Illinois, 1110 W.\ Green Street, 
Urbana, Illinois 61801, USA}
\author{Martin Weigel}
\affiliation{Applied Mathematics Research Centre, Coventry University, Coventry CV1 5FB, UK}
\author{Zohar Nussinov}
\email{zohar@wuphys.wustl.edu}
\affiliation{Department of Physics, Washington University, St. Louis, MO 63160, USA}

\begin{abstract}
  To establish a unified framework for studying both discrete and continuous coupling
  distributions, we introduce the {\it binomial} spin glass, a class of models where the
  couplings are sums of $m$ identically distributed Bernoulli random variables. In
  the continuum limit $m \to \infty$, the class reduces to one with Gaussian
  couplings, while $m=1$ corresponds to the $\pm J$ spin glass. We demonstrate that for
  short-range Ising models on $d$-dimensional hypercubic lattices the ground-state
  entropy density for $N$ spins is bounded from above by $(\sqrt{d/2m} + 1/N)
  \ln2$, and further show that the actual entropies follow the scaling behavior
  implied by this bound. We thus uncover a fundamental non-commutativity of the
  thermodynamic and continuous coupling limits that leads to the presence or absence
  of degeneracies depending on the precise way the limits are taken. Exact
  calculations of defect energies reveal a crossover length scale
  $L^\ast(m) \sim L^\kappa$ below which the binomial spin glass is indistinguishable
  from the Gaussian system. Since $\kappa = -1/(2\theta)$, where $\theta$ is the
  spin-stiffness exponent, discrete couplings become irrelevant at large scales for
  systems with a finite-temperature spin-glass phase.
 \end{abstract}

\maketitle

Spin glasses are extremely rich systems that have continued to surprise for many
decades \cite{book1,Hertz,book2,get_real,Stein-Complexity,Nishimori-Book,
  Nishimori-Ortiz, paw1,EA, Lukic, David-Scott, Binder,Parisi}. They represent
paradigmatic realizations of complexity that are abundant in nature and numerous
combinatorial optimization problems \cite{Francisco}. Abstractions of spin-glass
physics have led to new optimization algorithms and new insight into computational
complexity \cite{cavity,MPV,science,sp}, shed light on protein folding
\cite{protein}, and provided models of neural networks \cite{neuron}. Notwithstanding
this success, several fundamental questions still linger. These include
\cite{Newman-degeneracy} the character of the low-lying states and whether there are
many incongruent \cite{def:incongruent} ground states. It has long been known that
spin-glass systems with discrete couplings may rigorously exhibit an extensive
degeneracy \cite{Avron,loebl}, but these results do not extend to continuous coupling
distributions \cite{sue,Marinari, Houdayer, Rieger,Bhatt}. The possibility of
vanishing spectral gaps mandates the distinction of localized and extended
excitations, and only the latter can give rise to a multitude of states.
  
In this paper, we connect the $\pm J$ and the Gaussian spin glass models by
interpolating them via the {\em binomial\/} spin glass that has a tunable control
parameter $m$. We establish bounds of the spectral degeneracy of the Ising system on
bipartite graphs, which includes the usual Edwards-Anderson (EA) model with $\pm J$
($m=1$) and Gaussian ($m\to\infty$) couplings \cite{EA,Notem,rigor,talagrand,Mezard_SM,overlap,ultra,franz,two1,two2,two3,two4,two5,two6,two7,two8}. We thus show that
discrete (finite $m$) spin-glass samples exhibit an extensive ground-state
degeneracy, while continuous ones ($m\to\infty$) become two-fold degenerate, while
more generally the degeneracy depends on the precise way the non-commuting limits
$N\to\infty$ and $m\to\infty$ are taken.

We define the {\it binomial Ising spin glass} on a graph of $N$ sites
\cite{ourstopological} by the Hamiltonian
\begin{eqnarray}
\label{H_m}
H_m= - \sum_{\langle \sf{xy} \rangle}  {\cal J}^{m}_{\sf{xy}} s_{\sf{x}}s_{\sf{y}} \equiv 
- \sum_{\alpha=1}^{\cal L}  {\cal J}^{m}_\alpha z_\alpha.
\end{eqnarray}
Here, the sum is over sites $\sf x$ and $\sf y$, defining a link
$\alpha = \langle\sf x \sf y\rangle$, ${\cal L}$ denotes the total number of links,
and $s_{\sf x} = \pm 1$.  The {\it binomial coupling} for each link $\alpha$,
 ${\cal J}^{m}_\alpha \equiv \frac{1}{\sqrt{m}} \sum^{m}_{k=1} J^{(k)}_\alpha$,
 is a sum of $m$ copies (or ``layers'') of binary couplings $J^{(k)}_\alpha = \pm 1$,
 each with probability $p$ of being $+1$. The probability distribution of
 ${\cal J}^{m}_\alpha$,
\begin{eqnarray}\hspace*{-0.4cm}
  \label{ProbJm}
  \tilde{P}({\cal J}^{m}_\alpha)=\sum_{j=0}^m \binom{m}{j} p^{m-j}(1-p)^j \delta \!\left ( {\cal J}^{m}_\alpha - 
  \frac{m-2j}{\sqrt{m}}\right ),
\end{eqnarray}
is a binomial. In the large-$m$ limit, the distribution (\ref{ProbJm}) approaches a
Gaussian of mean $\sqrt{m}(2p-1)$ and variance $\sigma^{2} = 4p(1-p)$. In particular,
for $p=1/2$, the distribution $\tilde{P}({\cal J}^{m}_\alpha)$ approaches the
standard normal distribution usually considered for the EA model \cite{EA}.

To understand the degeneracies in the spectrum, we study the entropy density of the
$\ell$-th energy level,
\begin{eqnarray} \hspace*{-0.4cm}
  {\cal S}_{\ell}\equiv 
\frac{\sum  \limits_{\{  {\cal J}^{m}_\alpha \}} \! P(\{  {\cal J}^{m}_\alpha \})
  \ln D_{\ell} (\{  {\cal J}^{m}_\alpha \})}{N},
  \label{Entropy}
\end{eqnarray} 
where $D_{\ell}$ is the degeneracy of the $\ell$-th energy level \cite{Avron}.
$P(\{ {\cal J}^{m}_\alpha \})=\prod_{\alpha=1}^{\cal L} {\tilde P}({\cal
  J}^{m}_\alpha)$ is the probability of the coupling configuration.

We first embark on the derivation of an upper bound on the ground state entropy
density ${\cal S}_0$. We restrict ourselves to bipartite graphs, where any closed
loop encompasses an even number of links $\alpha$. Consider two spin configurations
$| {\sf{s}} \rangle \neq | {\sf{s'}} \rangle$ and evaluate their energy difference
$\Delta E=E({\sf s})-E({\sf s'})$.  From Eq. \eqref{H_m},
\begin{eqnarray} \hspace*{-0.4cm}
\label{DE}
\Delta E &=& - \sum^{\cal L}_{\alpha=1} {\cal J}^{m}_\alpha \Big(z_{\alpha}({\sf s}) - z_{\alpha}({\sf s'}) \Big) 
= - 2\sum^{\cal L}_{\alpha=1} {\cal J}^{m}_\alpha n_{\alpha},
\end{eqnarray}
with integers $n_{\alpha} = 0$, $\pm1$ defined by
$n_{\alpha} \equiv [z_{\alpha}({\sf s}) - z_{\alpha}({\sf s'})]/2$, where
$z_{\alpha}({\sf{s}}) = s_{\sf x} s_{\sf y}$. If $| {\sf{s}} \rangle$ and
$| {\sf{s'}} \rangle$ are degenerate then $\Delta E =0$. A degeneracy only occurs for
some realizations $\{{\cal J}^{m}_\alpha \}$ of the couplings, and Eq.~(\ref{DE}) can
be understood as a set of conditions for the couplings to ensure this.

Consider an arbitrary reference configuration $| {\sf{s}} \rangle$ of energy
$E(\sf{s})$ and examine its viable degeneracy with the contending $2^N -1$ other
configurations ${| {\sf{s'}} \rangle}$.  Each of these leads to a particular set of
integers ${\sf C}_{j}=\{n_{\alpha}\}_j$, which form the set
$\{{\sf C}_{j}\}_{j =1,2^N-1}^{| {\sf{s}} \rangle}$. A subset of those,
${\sf Sat}_{| {\sf{s}} \rangle}=\{{\sf C}_{j_1}, {\sf C}_{j_2}, \cdots, {\sf
  C}_{j_{\cal N}}\}$, will satisfy the degeneracy condition $\Delta E =0$ in
Eq.~\eqref{DE} for some coupling realizations. There are two types of solutions to
the equation $\Delta E =0$: (i) $n_{\alpha} = 0, \forall \alpha$, or (ii)
$n_{\alpha} \neq 0$, for at least one link $\alpha$.
It is straightforward to demonstrate that there is a single configuration
$| {\sf{s'}} \rangle ( \neq | {\sf{s}} \rangle$) for which {(i)}
$n_{\alpha} =0, \forall \alpha$ \cite{explain_triv*}. This is the degenerate
configuration $| {\sf{s}}' \rangle$ obtained by inverting all of the spins in
$| {\sf{s}} \rangle$. To determine whether the degeneracy may be larger than two, we
need to compute the probability ${\cal P}$ that constraints of type {(ii)} may be
satisfied.
While we cannot exactly calculate this probability
for general $N$ and $m$, bounds that we will derive suggest that
$\lim_{N \to \infty} \lim_{m \to \infty} {\cal S}_{\ell} =0$.
As we will emphasize, different large $m$ and $N$ limits may yield incompatible
results.

Constraints ${\sf{C}}_j\in {\sf Sat}_{| {\sf{s}} \rangle}$ are in {\it a one-to-one
  correspondence} with zero-energy interfaces \cite{explain_constraint*}, whose {\it
  size} is equal to the number $g_j$ of non-zero integers in the set
$\{n_{\alpha}\}_j$.  That is, given a fixed reference configuration
$| {\sf{s}} \rangle$ and a degenerate one $|{\sf s}' \rangle$, all type (ii)
solutions to Eq.~(\ref{DE}) are associated with configurations where the product
$s^{\;}_{\sf x} s'_{\sf x}$ is equal to $-1$ in a non-empty set of sites
${\sf{x}} \in R$.  To avoid the trivial redundancy due to global spin inversion,
consider the states $|{\sf s} \rangle$ and $|{\sf s}' \rangle$ for which the spin at
an arbitrarily chosen ``origin'' of the lattice assumes the value $+1$.  These states
are related via $ | {\sf s'} \rangle = U_{\sf s's} | {\sf s} \rangle$, where the {\it
  domain-wall operator} $U_{\sf s's}$ is the product of Pauli matrices that flip the
sign of the spins $s'_{\sf x}$ at the sites ${\sf x}$ where $|{\sf s} \rangle$ and
$| {\sf s'} \rangle$ differ.
Regions $R$ are bounded by zero-energy domain walls that are interfaces dual to the
links with $n_\alpha=\pm 1$, i.e., surrounding the areas $R$ where the spins  in 
$| {\sf{s}} \rangle$ an $| {\sf{s}'} \rangle$ have opposite orientation. Each
satisfied constraint
${\sf C}_{j} \in {\sf Sat}_{| {\sf{s}}
  \rangle}$ 
is associated with a state $|{\sf s'} \rangle=U_{{\sf s'} {\sf s}} |{\sf s}\rangle$
that is degenerate with $|{\sf s}\rangle$ for some coupling realization(s).

We next formalize the counting of {\it independent domain walls} or clusters of free
spins to arrive at an asymptotic bound on their number [Eq.~(\ref{scale-nm})].  This
will, in turn, provide a bound on the degeneracy.
We define a {\em complete} set of {\em independent} constraints 
${\overline{\sf Sat}}_{| {\sf{s}} \rangle} \subset {\sf Sat}_{| {\sf{s}} \rangle}$, of cardinality ${\cal M}$,  to
be composed of {\it all} constraints
${\sf{C}}_{\bar{\j}}\in {\sf Sat}_{| {\sf{s}} \rangle}$ that lead to {\em linearly
  independent equations} of the form of Eq.~(\ref{DE}),
$\Delta E= E({\sf s}) - E({\sf s}_{\bar{\j}})=0$, on the coupling constants
$\{J^m_\alpha\}$ \cite{explain_constraint*}.  All constraints in
${\sf Sat}_{| {\sf{s}} \rangle}$ are a consequence of the linearly independent subset
of constraints ${\overline{\sf Sat}}_{| {\sf{s}} \rangle}$. Each constraint 
${\sf{C}}_{\bar{\j}}\in {\sf Sat}_{| {\sf{s}} \rangle}$ is associated with a domain wall operator 
$U_{{\sf s}_{\bar{\j}} {\sf s}}$ that generates a degenerate state $| {\sf s}_{\bar{\j}} \rangle = 
U_{{\sf s}_{\bar{\j}} {\sf s}} |{\sf s} \rangle$.
If for a given coupling
realization $\{ {\cal J}^{m}_\alpha\}$ there are $M(\{ {\cal J}^{m}_\alpha\})\leq {\cal M}$ 
such independently satisfied constraints, 
 then the states 
 \begin{eqnarray}
 \label{binary}
 |\bar{n}_{1} \bar{n}_{2}\cdots \bar{n}_{M} \rangle  \equiv U^{\bar{n}_{1}}_{{\sf s}_{\bar{1}} {\sf s}}
  U^{\bar{n}_{2}}_{{\sf s}_{\bar{2}} {\sf s}} \cdots
   U^{\bar{n}_{M}}_{{\sf s}_{\bar{M}} {\sf s}} | {\sf s} \rangle,
\end{eqnarray}
($\bar{n}_{i}
=0,1$) will include all of the spin configurations degenerate with $|{\sf s}
\rangle$. Taking global spin inversion into account, the degeneracy of $| {\sf s}
\rangle$ is
\begin{eqnarray}
\label{D2}
D_{ \ell (| {\sf s} \rangle, \{{\cal J}^{m}_\alpha\})} \le 2^{M(\{  {\cal J}^{m}_\alpha\})+1},
\end{eqnarray} 
where, for a system defined by the coupling constants $\{{\cal J}^{m}_\alpha\}$, the
index $\ell(| {\sf s} \rangle, \{{\cal J}^{m}_\alpha\}))$ denotes the level $\ell$
the state $| {\sf s} \rangle$ belongs to. The set
$\{ |\bar{n}_{1} \bar{n}_{2} \cdots \bar{n}_{M} \rangle \}$ may contain additional
states not degenerate with $|{\sf s} \rangle$ \cite{independent_walls}.

After averaging over disorder, the expected number of the linearly independent
satisfied constraints ${\overline{\sf Sat}}_{| {\sf{s}} \rangle}$ is
\begin{eqnarray}\hspace*{-0.6cm}
\label{nm} 
\langle M  \rangle_{m} & \equiv& \hspace*{-0.1cm}
\sum_{\{ 
{\cal J}^{m}_{\alpha}\}} \sum_{{\sf C}_{\bar{\j}} \in {\overline{\sf Sat}}_{| {\sf{s}} \rangle}}  
\hspace*{-0.2cm} P(\{  {\cal J}^{m}_\alpha\})  \delta^{\{ {\cal J}^{m}_{\alpha}\}}( {\sf C}_{\bar{\j}}) 
\equiv \hspace*{-0.25cm} \sum_{{ \sf C}_{\bar{\j}} \in {\overline{\sf Sat}}_{| {\sf{s}} \rangle}} 
\hspace*{-0.3cm} {\cal P}({\sf C}_{{\bar{\j}} } ).
\end{eqnarray} 
Here, ${\cal P}({\sf C}_{{\bar{\j}} })$ is the probability that a linearly independent
constraint ${\sf C}_{{\bar{\j}} }$ is satisfied. The Kronecker
$\delta^{\{ {\cal J}^{m}_{\alpha}\}}( {\sf C}_{\bar{\j}})$ equals $1$ if
$ {\sf C}_{\bar{\j}}$ is satisfied for the couplings $\{ {\cal J}^{m}_{\alpha}\}$ and
is zero otherwise. Let us bound the probability ${\cal P}({\sf C}_{{\bar{\j}} })$ by
taking the form (\ref{ProbJm}) of the coupling distribution into account. 
From the definition of the couplings
$\{{\cal J}^{m}_\alpha\}$,  the sum in Eq.~\eqref{DE} can effectively
be read as including a sum over layers $k=1,\ldots,m$, which hence includes
$g_{\bar{\j}} m$ non-zero terms. For general $m\geq 1$, and even $g_{\bar{\j}} m$, the
probability that half of the nonzero integers $n_{\alpha} J^{(k)}_\alpha$ in
Eq.~\eqref{DE} are $+1$ and the remainder are $-1$ is
\begin{eqnarray}
\label{m-scale}
{\cal P} ({\sf C}_{{\bar{\j}} }) = 
\binom{g_{\bar{\j}}m}{\frac{g_{\bar{\j}}m}{2}} \frac{1}{2^{g_{\bar{\j}}m}} < \frac{1}{ \sqrt{g_{\bar{\j}} m}}.
\end{eqnarray}
(Eq.~(\ref{DE}) cannot be satisfied for odd $g_{\bar{\j}} m$.) From asymptotic
analysis \cite{Note1} and Eq.~(\ref{m-scale}), the probability
${\cal P}({\sf C}_{{\bar{\j}} })$ scales (for large $m$) as (and, for any $m$, is
bounded by) $1/\sqrt{g_{\bar{\j}} m}$.  Denoting by $g_{\min}$ the smallest possible
value of $g_{\bar{\j}}$ for the graph/lattice at hand,
\begin{eqnarray}
\label{scale-nm}
\langle M \rangle_{m} \le \frac{{\cal M}}{\sqrt{g_{\min}m}}.
\end{eqnarray}
On a general graph, the number ${\cal M}$ of linearly {\it independent} constraints 
${\sf C}_{\bar{\j}}$ on the coupling constants $\{{\cal J}^{m}_{\alpha}\}$ 
cannot be larger than their total number, ${\cal{M}} \le {\cal L}$,
i.e., the number of links ${\cal L}$ on this graph.
Putting all of the pieces together, Eqs.~(\ref{D2}) and (\ref{scale-nm}) imply
\begin{eqnarray} \hspace*{-0.7cm}
\label{PJ}
\sum_{\{{\cal J}^{m}_\alpha\}} P (\{{\cal J}^{m}_\alpha\}) \ln D_{ \ell (| {\sf s} \rangle, \{{\cal J}^{m}_\alpha\})} 
\le (1+ \frac{{\cal L}}{\sqrt{ g_{\min} m}}) \ln 2.
\end{eqnarray} 
Trying to evaluate the l.h.s.\ of Eq.~(\ref{PJ}) we must take into account that
whatever $| {\sf s} \rangle$ we pick might be a ground state for some coupling
configurations, but will be an excited state for others. Hence we cannot directly
infer a bound to the average entropy ${\cal S}_\ell$ from (\ref{PJ}). Since the
inverse temperature $1/(k_B T)=\partial \ln D/\partial E$, however, the system's
ground-state degeneracy for couplings $\{{\cal J}^{m}_\alpha\}$ is typically lower
than (or equal to) that of any other level $\ell$ \cite{random_lev}, i.e.,
$D_0\leq D_{\ell}$. This monotonicity of $D(E)$ implies that, typically,
${\cal S}_0 N=\sum_{\{{\cal J}^{m}_\alpha\}} P (\{{\cal J}^{m}_\alpha\}) \ln D_0(\{{\cal J}^{m}_\alpha\})   \le 
\sum_{\{{\cal J}^{m}_\alpha\}} P (\{{\cal J}^{m}_\alpha\}) 
\ln D_{ \ell (| {\sf s} \rangle,\{{\cal J}^{m}_\alpha\})}$. Then, 
Eq.~(\ref{PJ}) yields 
\begin{eqnarray}
\label{final!}
{\cal{S}}_{0} \le (\frac{{\cal L}}{N\sqrt{g_{\min} m}} + \frac{1}{N}) \ln 2.
\end{eqnarray}
This is the promised rigorous bound. For $p \neq 1/2$ one has a lower entropy density
than that of $p=1/2$. Thus, Eq.~(\ref{final!}) constitutes a generous upper bound on
${\cal S}_{0}$ for general $p$. To study higher energy levels, consider the average
of Eq.~(\ref{PJ}) over all possible $2^{N}$ reference spin configurations
$| {\sf s} \rangle$. Performing this average and invoking the monotonicity of $D(E)$
suggests that the entropy density ${\cal S}_{\ell}$ of Eq.~(\ref{Entropy}) of
low-lying excited levels $\ell>0$ is, typically, also bounded by the r.h.s\ of
Eq.~(\ref{final!}). For $d$-dimensional hypercubic lattices with periodic boundary
conditions, the ratio ${\cal L}/N =d$ while $g_{\min} = 2d$. Thus,
$ {\cal S}_{0} \le (\sqrt{d/2m} + 1/N) \ln2$.  Eq.~(\ref{final!}) further suggests
that, in the thermodynamic ($N \to \infty$) limit \cite{explain_as*},
 \begin{eqnarray}
\label{final-conjecture}
{\cal S}_{0}(m') \sim \sqrt{\frac{m}{m'}} \, {\cal S}_{0}(m)  \ ~ \ \mbox{for~finite ~$m,m' \gg 1$}.
\end{eqnarray}

\begin{figure}
  \includegraphics[width= \linewidth]{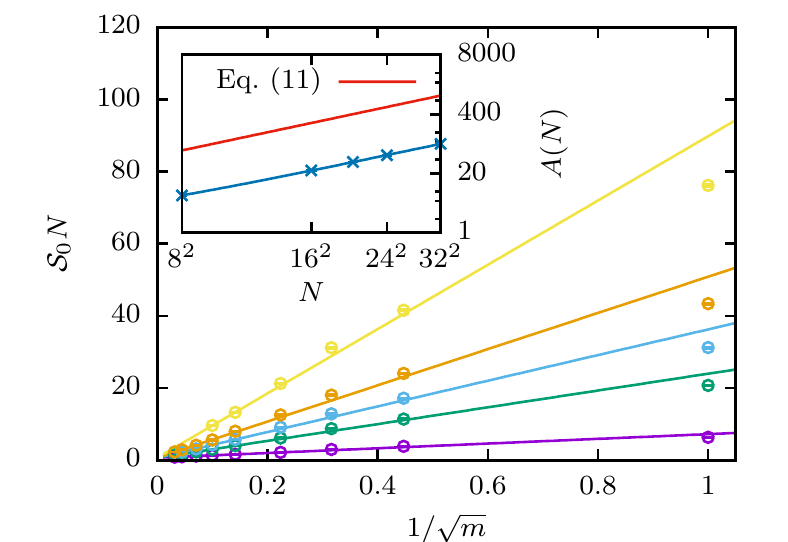}
  \caption{Ground-state entropy ${\cal S}_0N$ of the binomial Ising spin glass with
    $m$ layers, cf.\ Eq.~(\ref{H_m}), on square lattices of $N=L^2$ spins from exact
    ground-state calculations (from the bottom: $L=8$, $16$, $20$, $24$, and $32$).
    Lines are fits of the form of (\ref{eq:entropy-scaling}) to the data for
    sufficiently large $m$. The inset shows the linear scaling of the amplitude
    $A(N)$. The top line indicates the constraint imposed by the upper bound
    \eqref{final!}.}
  \label{fig:lscaling}
\end{figure}

We now study the exact $m$ dependence of the ground state entropies of the binomial
model on the square lattice with periodic boundaries and $N=L^2$. To this end, we employed an
implementation of the Pfaffian technique of counting dimer coverings of the lattice
as discussed in Ref.~\cite{vondrak}, which is a generalization of earlier methods
\cite{blackman,kardar} to fully periodic lattices. In Fig.~\ref{fig:lscaling}, we
present the results for the ground-state entropy, averaged over 1000 coupling
realizations for each lattice size. The data are well described by
\begin{equation}
  {\cal S}_0 N = \left(\frac{A(N)}{\sqrt{m}} + 1\right)\ln 2.
  \label{eq:entropy-scaling}
\end{equation}
Linear fits in $1/\sqrt{m}$ for fixed $N$ work well for sufficiently large $m$, as is
illustrated by the straight lines in Fig.~\ref{fig:lscaling}. Thus, for any finite
$N$, as $m\to\infty$ the ground-state entropy is equal to $\ln 2$, implying a single
degenerate ground-state pair. The slope $A(N)$ shown in the inset follows a linear
behavior, $A(N) = a N + b$, and we find $a = 0.0858(4)$ and $b=1.09(12)$. For not too
small $m$, our data are hence fully consistent with
\begin{equation}
\label{snumeric}
  {\cal S}_0 = \left(\frac{a}{\sqrt{m}} + \frac{1}{N} + \frac{b}{N\sqrt{m}}\right)\ln 2.
\end{equation}
When $N \gg \sqrt{m} \gg 1$, Eq.~(\ref{snumeric}) is consistent with the physically
inspired \cite{explain_as*} scaling of Eq.~(\ref{final-conjecture}). For large $N$,
the bound of Eq.~\eqref{final!} would have been asymptotically saturated if
$a \simeq 1$, far larger than the actual value of $a$. The behavior in the double
limit $m ,N \to \infty$ is subtle: (1) for $m \to \infty$, $N$ finite, we have a
single ground-state pair; (2) for $N \to \infty$, $m$ finite, there is a finite
ground-state entropy $\sim \ln 2/\sqrt{m}$; (3) for $N \to \infty$, $m \to \infty$,
$\kappa = N/\sqrt{m}$ fixed, there is a finite number $2^{a\kappa}$ of ground-state
pairs. Thus clearly the continuum and thermodynamic limits are not commutative in
general.  Note further that according to the bound
${\cal S}_{0} \le (\sqrt{d/2m} + 1/N) \ln2$ for hypercubic lattices additional rich
behavior is expected if the limit of high dimensions is correlated with that of large
$m$.

\begin{figure}
  \includegraphics[width=\linewidth]{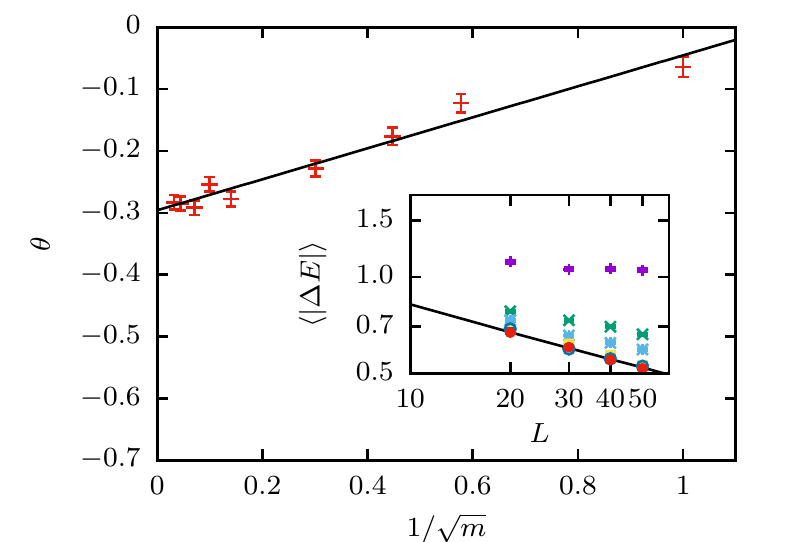}
  \caption{ Effective spin stiffness exponents $\theta = \theta(m)$ resulting from
    fits of the power law $\langle |\Delta E|\rangle = B L^\theta$ to the defect
    energies for the binomial model of $m$ layers (inset, from the
    top: $m=1$, $5$, $11$, $51$, $201$, and $1001$), averaged over $10\,000$ disorder
    samples. The solid line of the inset corresponds to the Gaussian model.}
  \label{fig:defects}
\end{figure}

\begin{figure}
  \includegraphics[width=\linewidth]{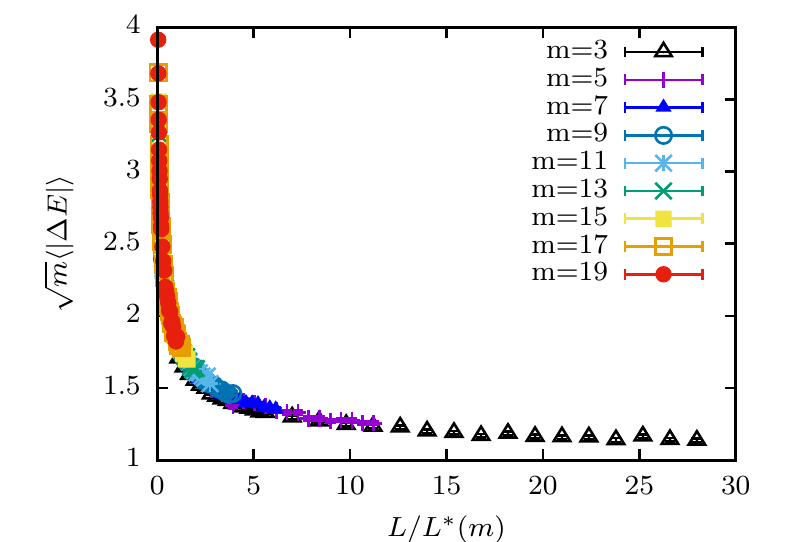}
  \caption{ Scaling collapse of the defect energies of the binomial model for system
    sizes rescaled with the crossover length scale $L^\ast(m) \sim m^\kappa$ with
    $\kappa = 1.79$.  }
  \label{fig:crossover}
\end{figure}

Let us turn to the study of excitations.  By construction, cf.\ Eq.~(\ref{DE}), for
finite $m$ the energy is ``quantized'' in multiples of $1/\sqrt{m}$. It is therefore
natural to expect a closing of the spectral gap as $m\to\infty$. That this is indeed
the case can be shown rigorously for the one-dimensional binomial spin glass in its
thermodynamic limit, with different behaviors for odd and even $m$, see the
discussion in the Supplemental Material \cite{Note4}. The closing of the gap is a
consequence of the existence of (rare) local excitations, i.e., finite-size clusters
of almost free spins \cite{Note3}.  Whether gapless {\em non-local\/} excitations
exist and which form they take in the thermodynamic limit is a long-standing question
\cite{kawashima:03a}. One possible approach of investigating such excitations
consists of subjecting individual samples to a system spanning perturbation by a
change of boundary condition and studying how this affects the energy and
configuration of the ground state. Such defect energy calculations \cite{cieplak:83}
enable us to extract a scaling $\langle |\Delta E|\rangle \sim L^\theta$ of the
defect energies with the spin stiffness exponent $\theta$. Generalizing Peierls'
argument \cite{Peierls1,Perierls2,clock,Bonati} for the stability of the ordered
phase, one should find $\theta > 0$ for cases where there is a finite-temperature
spin-glass phase, and $\theta \le 0$ otherwise. The latter case is expected for
dimensions $d=1$ and $d=2$, whereas $\theta$ is positive for $d \ge 3$
\cite{hartmann:01a,boettcher:04}. We employed techniques based on minimum-weight
perfect matching \cite{bieche:80a,khoshbakht:17a} to perform such calculations for
the binomial model on the square lattice. The resulting disorder-averaged defect
energies from {\it exact} ground-state calculations for samples with periodic and
antiperiodic boundaries are shown in the inset of Fig.~\ref{fig:defects}. As $m$
increases, the decay of defect energies as a function of $L$ becomes steeper and the
data approach the behavior of the Gaussian EA model. The effective spin stiffness
exponents $\theta$ extracted from fits of the type
$\langle |\Delta E|\rangle = B L^\theta$ are shown in the main panel of
Fig.~\ref{fig:defects}. These exponents appear to interpolate smoothly between the
limiting cases of the Gaussian model with $\theta = -0.2793(3)$ and the $\pm J$
system with $\theta = 0$ \cite{hartmann:01a,khoshbakht:17a}. Asymptotically, however,
we expect that $\theta(m) = 0$ for any finite value of $m$ when
$L \gtrsim L^\ast(m)$.  The scaling of the crossover length $L^\ast(m) \sim m^\kappa$
follows by considering the model with the unscaled couplings
$\sqrt{m} {\cal J}_\alpha^m$, for which the energy gap $\Delta$ is independent of
$m$. The discreteness of the spectrum becomes apparent once the corresponding defect
energies $\sqrt{m} \langle |\Delta E|\rangle \sim L^\theta$ have decayed below the
size of the gap, i.e., for
\[
  L \ge L^\ast(m) \sim m^{-1/(2\theta)},
\]
such that $\kappa = -1/(2\theta)$. For the $d=2$ system we have $\theta=-0.2793(3)$
\cite{khoshbakht:17a}, such that $\kappa = 1.790(2)$, which is in excellent agreement
with the actual defect energies for our system shown in Fig.~\ref{fig:crossover}.

It is clear that if $\theta<0$, as is the case for the Gaussian spin glass in two
dimensions, excitations of a divergent length scale may entail a vanishing energy
penalty. At zero temperature, the discreteness of the spectrum is then always seen at
large scales $L \gtrsim L^\ast(m)$. On the other hand, for $\theta \ge 0$ (i.e., $d \ge 3$), the above
arguments imply that the discreteness does not matter at large scales. Also, in this
case one should inspect the full probability distribution of domain wall energies and
the weight it carries in the limit $\Delta E \to 0$ \cite{Note3}. In how far such
excitations correspond to incongruent states, however, one might only be able to
infer by inspecting the configurations themselves.

In summary, we introduced and discussed the {\it binomial spin glass}. This class of models
affords controlled access to the enigmatic continuous ($m \to \infty$) finite
dimensional EA model. Its $m=1$ realization is the quintessential discrete spin
glass, the $\pm J$ model. We derived bounds on the spectral degeneracy of the
binomial Ising spin glass on general graphs and suggested an asymptotic scaling that is
fully supported by exact two-dimensional calculations.  The behavior of defect
energies suggests the existence of a crossover length $L^\ast(m) \sim L^{-1/2\theta}$
below which the binomial model behaves like the Gaussian system. Our results show
that the existence of degeneracies depends on the particular way of taking the
thermodynamic ($N \to \infty$) and continuous coupling $(m \to \infty$) limits, and
limiting states with and without degeneracies can be reached by corresponding
correlated limiting processes, thus accommodating theories that postulate
degeneracies as well as pictures stipulating a unique ground-state pair. An
intriguing prediction regards an effectively negative crossover scaling exponent in
three dimensions, where hence discreteness of the spectrum is expected not to matter
at large scales.

The physics of spin-glass models and, in particular, the role of degeneracies has
also recently attracted attention from another side. In the context of quantum
annealing \cite{finnila:94} as implemented in the devices by D-Wave and similar
machines that are being developed by competing consortia, degeneracies are not a
desired feature as the quantum annealing process does not sample such states
uniformly \cite{mandra:17}. On the other hand, continuous coupling distributions may
also be undesired because of increased susceptibility to external noise implied by
chaos in spin glasses \cite{bray:87,Thomas2011,dandan:12,zhu:16}. Our binomial
glasses may allow for realizations that suffer the least from these combined
problems. While the present system is already a generalization of the usually
considered spin-glass models, we believe that the approach of decomposing continuous
couplings into discrete layers and the intriguing consequences it allowed us to
uncover in terms of the general non-commutativity of the thermodynamic and continuous
coupling limits is promising and we expect exciting applications to  models in other
fields.

{\it{Acknowledgements}}. This research was partially supported by the NSF CMMT under
grant number 1411229.  

\newpage
\pagebreak

\newpage
\onecolumngrid
\pagebreak

\newpage

\renewcommand{\thefigure}{S\arabic{figure}}
\renewcommand{\theequation}{S\arabic{equation}} 
\setcounter{figure}{0}
\setcounter{equation}{0}
\setcounter{page}{1}

\begin{center}
{\large \bf Supplemental Material for The Binomial Spin Glass}\\
\vspace{0.4cm}

M.-S. Vaezi,$^1$ G. Ortiz,$^{2,3}$ M. Weigel,$^4$ and Z. Nussinov$^{1,*}$

{\small \it $^1$ Department of Physics, Washington University, St.
Louis, MO 63160, USA}

{\small \it $^2$ Department of Physics, Indiana University, Bloomington,
IN 47405, USA}

{\small \it $^3$ Department of Physics, University of Illinois, 1110 W. Green Street, 
Urbana, Illinois 61801, USA}

{\small \it $^4$ Applied Mathematics Research Centre, Coventry University, Coventry CV1 5FB, UK}

\end{center}
\vspace{0.4cm}

\twocolumngrid

Below, we further provide a lightning overview of the problem that prompted the current investigation (Section \ref{motivate}). We then elaborate on several aspects that were alluded to in the main text (Sections (B-H)).  

\subsection{General Background and Motivation}
\label{motivate}

The quintessential short-range Ising spin glass system is the Edwards-Anderson (EA)
model, where at each lattice site ${\sf{x}}$ lies a classical spin
$s_{\sf{x}} = \pm 1$, that interacts with nearest-neighbor spins only \cite{EA_SM}.  In
the {\it discrete} binary version, the random couplings may assume only the two
values $\pm J$. Conversely, the couplings are {\it continuous} random Gaussian
variables in the {\it continuous} EA model.  While the extensive ground state
degeneracy is well established for various binary distributions, the situation for
the continuous EA model has been mired by controversy.  Parisi's tour de force
solution \cite{Parisi_SM} led to insights concerning the extensive nature of the ground
state entropy of the infinite-range Sherrington-Kirkpatrick (SK) model
\cite{David-Scott_SM}.  The latter harbors a plethora of distinct thermodynamic states
\cite{book1_SM,rigor,talagrand}. A measure of similarity between disparate thermodynamic
states is provided by the well-known ``overlap function'' \cite{book1_SM,Mezard_SM,overlap}
$q_{r r'} = \frac{1}{N} \sum_{\sf{x}} \langle s_{\sf{x}} \rangle_{r} \langle
s_{\sf{x}} \rangle_{r'}$, where $N$ is the total number of lattice sites, and its
average over the probabilities $W_{r}$ and $W_{r'}$ of the realizations of the
different pairs of states $r$ and $r'$ (the ``overlap distribution function''),
$P(q) = \sum_{rr'} q_{rr'} W_{r} W_{r'}$.  The SK model displays a cascade of
different overlaps (an ultrametric structure \cite{ultra}) and replica symmetry
breaking wherein $P(q)$ becomes nontrivial \cite{franz}.  Standard ordered systems
typically display a small number of symmetry related thermodynamic states (and zero
temperature ground states) associated with a distribution $P(q)$ that is a sum of
simple delta functions.  While the Parisi solution and various related (effective
infinite dimension or infinite range) mean-field treatments raise the possibility of
an exponentially large number of ground states, other considerations
\cite{book1_SM,two1,two2,two3,two4,two5,two6,two7,two8} suggest that (similar to ferromagnets)  in typical
short-range spin glasses, there are only two symmetry
related ground states. The understanding of this problem underlies our work. This question is not merely of academic
importance; the behavior of real finite dimensional magnetic spin glass systems has
long been of direct experimental pertinence, e.g., \cite{book2_SM,get_real_SM}.


We now explicitly define the standard EA model.
Consider a general bipartite lattice (in any finite number of dimensions $d$) of size 
$N$, endowed with periodic boundaries, 
with an Ising spin $s_{\sf x}$ at each lattice site $\sf x$. The EA spin glass Hamiltonian is given by
\begin{eqnarray}
\label{H_0}
H= - \sum_{\langle \sf{xy} \rangle} J_{\sf{xy}}s_{\sf{x}}s_{\sf{y}} \equiv - \sum_{\alpha=1}^{\cal L} J_{\alpha}z_\alpha.
\end{eqnarray}
The summation in Eq. (\ref{H_0}) is over nearest-neighbor spins at sites ${\sf{x}}$ and ${\sf{y}}$ sharing the link 
$\alpha = \langle {\sf xy} \rangle$, $z_{\alpha}=\pm 1$, and the total number of these links is  ${\cal L} = d\times N$.  
In various standard Ising spin glass models, the spin couplings 
 $\{J_\alpha \} $ in Eq. \eqref{H_0} are customarily drawn from one of several well studied distributions. 
For instance, in the ``binary Ising 
spin glass model'' \cite{Lukic_SM}, the couplings $\{J_\alpha \} $ are random variables that assume the two values $\pm1$ with 
probabilities $P(J_\alpha  =1) = p$, $P(  J_\alpha  =-1) = 1-p$ (i.e., a Bernoulli distribution). 
 In the continuous  EA model  the couplings $\{J_\alpha\}$ are drawn from a 
Gaussian distribution of vanishing mean and  variance equals to unity.

\subsection{The trivial ground state pair given an assignment of link variables}
\label{trivpair}

Given the definition of the link variable $z_{\alpha} \equiv s_{\sf{x}}s_{\sf{y}}$, a moment's reflection reveals that
\begin{eqnarray}
s_{\sf{y}} = s_{\sf{x}} \prod_{\alpha \in \Gamma_{\sf{xy}}} z_{\alpha} ,
\end{eqnarray}
where $\Gamma_{\sf{xy}}$ is any path on the lattice, composed of nearest-neighbor links, 
joining site $\sf{x}$ to site $\sf{y}$.
Thus, with $s_{\sf{y}}|_{|{\sf{s} \rangle}}$ denoting the value of the spin at site ${\sf{y}}$ in configuration $| {\sf{s}} \rangle$, we have that
\begin{eqnarray} \hspace*{-0.5cm}
\label{C1_C2}
s_{\sf{y}}|_{| {\sf{s}} \rangle} = s_{\sf{x}}|_{| {\sf{s}} \rangle} \prod_{\alpha \in \Gamma_{\sf{xy}}} z_{\alpha}|_{| {\sf{s}} \rangle} \ , 
s_{\sf{y}}|_{| {\sf{s'}} \rangle} = s_{\sf{x}}|_{{| {\sf{s'}}} \rangle} \prod_{\alpha \in \Gamma_{\sf{xy}}} z_{\alpha}|_{| {\sf{s'}} \rangle}.
\end{eqnarray}
Now, if for all links $\alpha$, the values of $z_\alpha$ are the same in both configurations 
${| {\sf{s}} \rangle}$ and ${| {\sf{s'}} \rangle}$ (i.e., if $\{z_{\alpha}\}|_{| {\sf{s}} \rangle} = \{z_{\alpha}\}|_{| {\sf{s'}} \rangle}$) then, trivially,
\begin{eqnarray}
\label{Prod}
\prod_{\alpha \in \Gamma_{\sf{xy}}} z_{\alpha}|_{| {\sf{s}} \rangle} = \prod_{\alpha \in \Gamma_{\sf{xy}}} z_{\alpha}|_{| {\sf{s'}} \rangle}.
\end{eqnarray}
Taken together, Eqs.~\eqref{C1_C2} and \eqref{Prod} imply that if, at a particular site ${\sf x}$, the  spin 
configurations $| {\sf{s}} \rangle$ and $| {\sf{s'}} \rangle$ share the same value of the spin, 
$s_{\sf{x}}|_{| {\sf{s'}} \rangle} = s_{\sf{x}}|_{| {\sf{s}} \rangle}$, then 
the spins must be identical at all other lattices sites ${\sf y}$, $s_{\sf{y}}|_{| {\sf{s'}} \rangle} = s_{\sf{y}}|_{| {\sf{s}} \rangle}$. This, however,  
leads to a contradiction as $| {\sf{s'}} \rangle \neq | {\sf{s}} \rangle$. Therefore, if two distinct spin configurations satisfy condition ({\bf{i}})  
it must be that the respective spin values at any lattice site ${\sf x}$ are different, $s_{\sf{x}}|_{| {\sf{s'}} \rangle} = - s_{\sf{x}}|_{| {\sf{s}} \rangle}$. 
That is,
\begin{eqnarray}
s_{\sf{y}}|_{| {\sf{s'}} \rangle} = -s_{\sf{y}}|_{| {\sf{s}} \rangle},~~~\forall {\sf{y}}.
\end{eqnarray}
Hence, if $n_{\alpha} =0,\forall \alpha$ in Eq. \eqref{DE} of the main text, then there are, trivially, only two degenerate configurations 
($| {\sf{s'}} \rangle \neq | {\sf{s}} \rangle$) related by a global spin inversion.
The above simple proof applies for arbitrary energy levels. Replicating, {\it mutatis mutandis}, the above argument to 
a general set of (non-necessarily vanishing) integers $\{n_{\alpha}\}$ over all lattice links $\alpha$,  illustrates that any set 
$\{n_{\alpha}\}$ may correspond to exactly two unique spin configurations.

\subsection{Graphical Representation of the Constraints}
\label{GRC}

In the main text we defined ${\sf Sat}_{| {\sf{s}} \rangle}$ to be the set composed of {\it all} 
constraints ${\sf{C}}_{j}$
satisfying the relation $\Delta E= E({\sf s}) - E({\sf s'})=0$, in Eq.~(\ref{DE}) of the main text.
We also defined the subset ${\overline{\sf Sat}}_{| {\sf{s}} \rangle} \subset {\sf Sat}_{| {\sf{s}} \rangle}$,
comprising {\it all linearly independent} constraints.
Here, we further introduce a restricted subset 
of constraints, that of  {\it geometrically disjoint} and independent zero energy domain walls, 
 ${\overline{\sf Sat^{g}}}_{| {\sf{s}} \rangle} \subset
 {\overline{\sf Sat}}_{| {\sf{s}} \rangle}$. The subset ${\overline{\sf Sat^{g}}}_{| {\sf{s}} \rangle}$ is defined by having no 
 pair of different constraints on the coupling constants that 
 involve links associated with the same lattice sites ${\sf x}$.

\begin{figure}
  \includegraphics[width= \columnwidth]{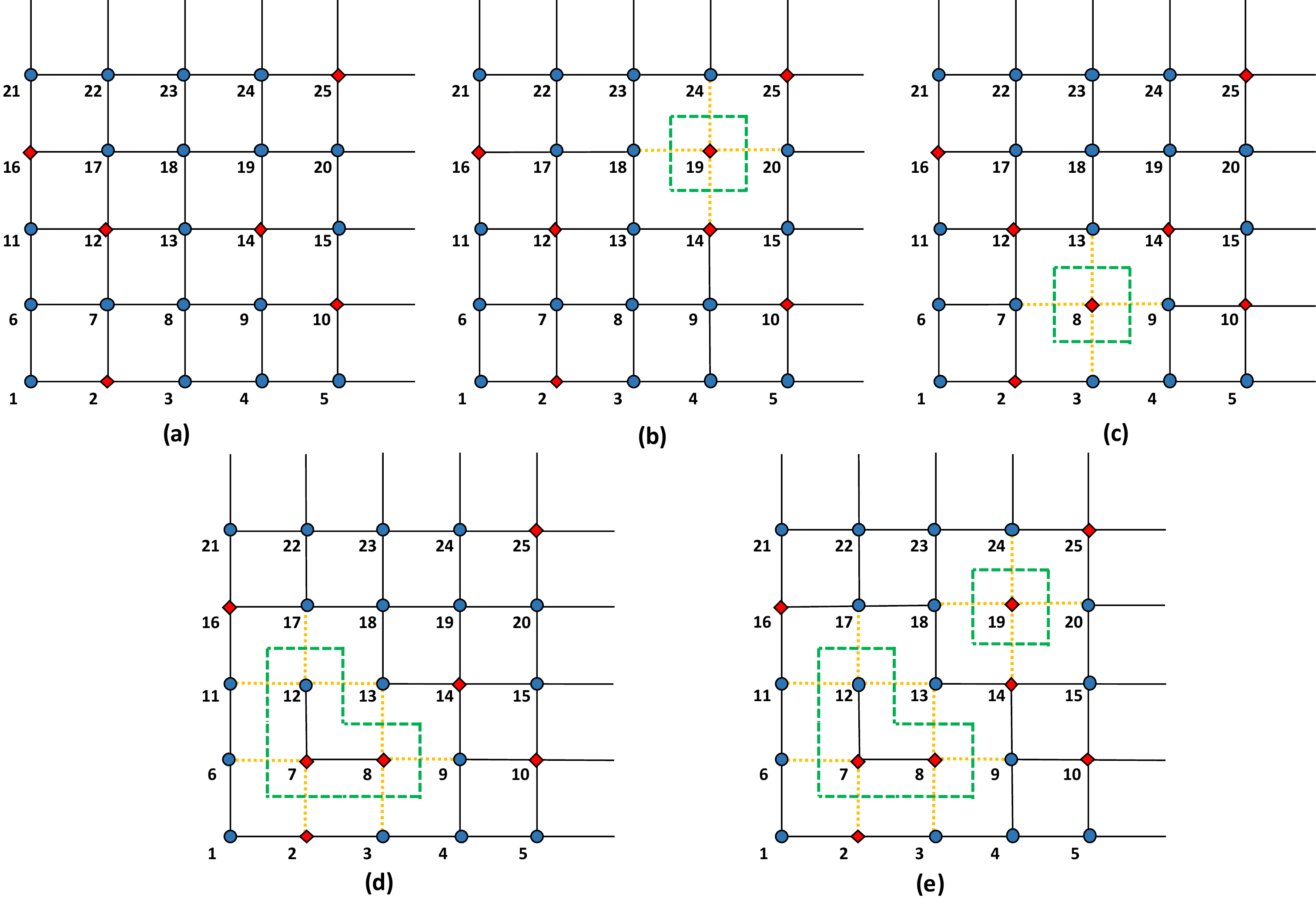}
  \caption{ Graphical representations of the constraints. 
  Panel (a) represents a random spin configuration.
  Blue solid circles and red diamonds denote spin up and down, respectively. 
  Flipping one or more spins at different sites of panel (a) would result in new spin configurations such as in panels (b) through (e)
  (e.g., the spin configuration of panel (b) is obtained from flipping the spin at site $19$ of panel (a)). 
  The dashed yellow dotted lines represent the links that contribute to the energy difference.
  The green dashed lines crossing such links correspond to a domain wall. }
  \label{fig:GRC_1}
\end{figure}

In what follows, we provide a few simple examples illuminating the above definitions. 
To this end, we consider a $5\times 5$ square lattice with binomial couplings 
$\{{\cal J}^m_\alpha\}$ (Fig.~\ref{fig:GRC_1}).
We start with a random spin configuration $|{\sf s} \rangle$ (panel (a)).
Panels (b) through (e), represent spin configurations $|{\sf s'} \rangle$ for which one or more 
spins are being flipped with respect to panel (a).
The energy difference in each case can be easily calculated. For example,
\begin{eqnarray}
\Delta E_{a,b} &=& E_a - E_b = -2({\cal J}^{m}_{19,14} n_{19,14} \nonumber\\
&+& {\cal J}^{m}_{19,18} n_{19,18} + {\cal J}^{m}_{19,20} n_{19,20} + {\cal J}^{m}_{19,24} n_{19,24}),\nonumber\\
\end{eqnarray}
gives the energy difference between spin configurations in panel (a) and (b). 
It is easy to see that $n_{19,18} = n_{19,20} = n_{19,24} = 1$, and $n_{19,14} = -1$.
Following the same procedure we end up with,
\begin{eqnarray}
\label{Energy_dif_SM}
\Delta E_{a,b} &=& -2(-{\cal J}^{m}_{19,14} + {\cal J}^{m}_{19,18} + {\cal J}^{m}_{19,20} +{\cal J}^{m}_{19,24}),\nonumber\\
\Delta E_{a,c} &=& -2({\cal J}^{m}_{8,3} + {\cal J}^{m}_{8,7} + {\cal J}^{m}_{8,9} + {\cal J}^{m}_{8,13}),\nonumber\\
\Delta E_{a,d} &=& -2(-{\cal J}^{m}_{7,2} + {\cal J}^{m}_{7,6} + {\cal J}^{m}_{8,3} + {\cal J}^{m}_{8,9} + {\cal J}^{m}_{8,13} \nonumber\\
&&- {\cal J}^{m}_{12,11} - {\cal J}^{m}_{12,13} - {\cal J}^{m}_{12,17}),\nonumber\\
\Delta E_{a,e} &=& \Delta E_{a,b} + \Delta E_{a,d} \nonumber\\
&=& -2(-{\cal J}^{m}_{19,14} + {\cal J}^{m}_{19,18} + {\cal J}^{m}_{19,20} + {\cal J}^{m}_{19,24} \nonumber\\
&&- {\cal J}^{m}_{7,2} + {\cal J}^{m}_{7,6} + {\cal J}^{m}_{8,3} + {\cal J}^{m}_{8,9} + {\cal J}^{m}_{8,13} \nonumber\\
&& -{\cal J}^{m}_{12,11} - {\cal J}^{m}_{12,13} - {\cal J}^{m}_{12,17}).\nonumber\\
\end{eqnarray}
Now, assume ${\sf C_1},{\sf C_2},{\sf C_3}$, and ${\sf C_4}$ are constraints associated with 
$\Delta E_{a,b}, \Delta E_{a,c}, \Delta E_{a,d}, \Delta E_{a,e}$, respectively. If these 
constraints are satisfied, i.e., $\Delta E_{a,,b}=\Delta E_{a,c}=\Delta E_{a,d}= \Delta E_{a,e}=0$, for certain 
coupling realizations, then they belong to the set  ${\sf Sat}_{| {\sf{s}} \rangle}$. That is, 
${\sf C_1}, {\sf C_2},{\sf C_3},{\sf C_4} \in  {\sf Sat}_{| {\sf{s}} \rangle}$.

\begin{figure}[htb]
  \includegraphics[width= \columnwidth]{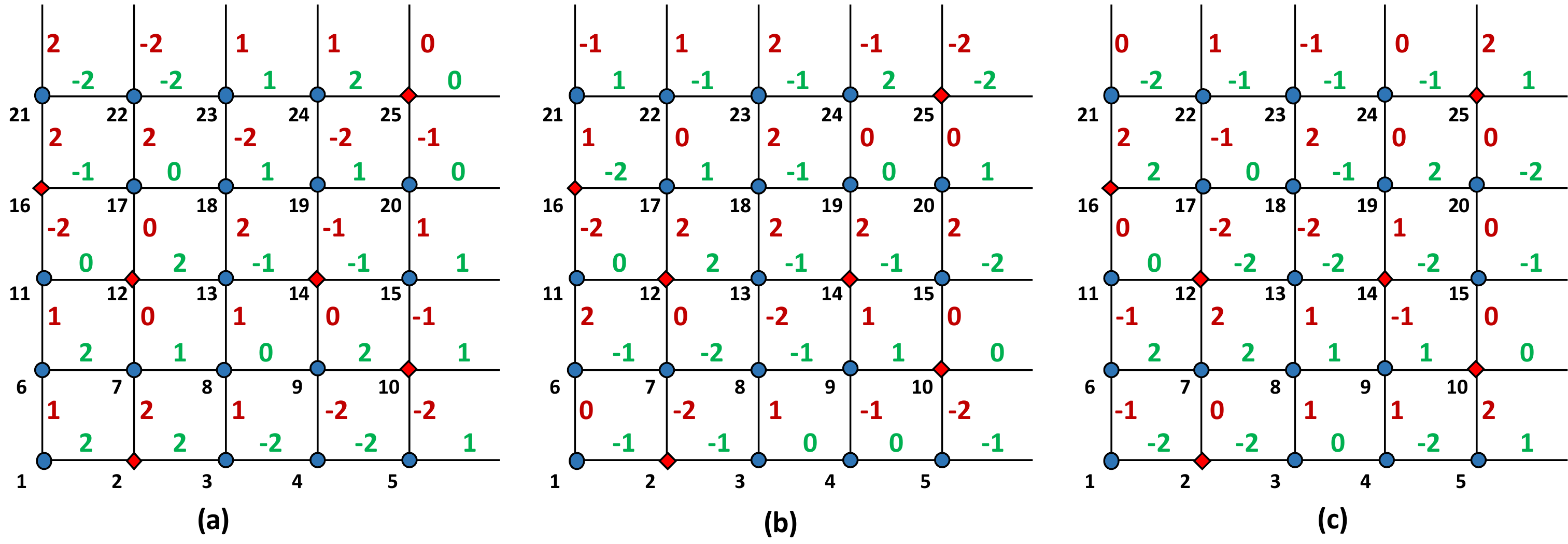}
  \caption{ Three examples of coupling realizations for the binomial model with $m = 4$ (i.e., ${\cal J}^4_{\alpha} = -2,-1,0,1,2$).
  The numbers in green (brown) color provide the values of horizontal (vertical) coupling constants.}
  \label{fig:GRC_2}
\end{figure}

To understand this better, consider the case $m = 4$.
Since ${\cal J}^{m}_\alpha \equiv \frac{1}{\sqrt{m}} \sum^{m}_{k=1} J^{(k)}_\alpha$, the couplings ${\cal J}^4_{\alpha}$ may acquire the values $-2,-1,0,1,2$.
In Fig.~\ref{fig:GRC_2}, we provide three examples of random coupling realizations.
The spin configuration is the same as in panel (a) of Fig.~ \ref{fig:GRC_1}.
From Eq. \eqref{Energy_dif_SM} and Fig.~\ref{fig:GRC_2}, we can see that, only ${\sf C_3}$ in panel (a), 
none in panel (b), and only ${\sf C_1}$ in panel (c) are satisfied.

In order to create the subset  ${\overline{\sf Sat}}_{| {\sf{s}} \rangle}$, we should note that it is not necessarily unique,
since we may have different linearly independent constraints that span the same set of conditions in ${\sf Sat}_{| {\sf{s}} \rangle}$.
In addition to that, the satisfaction of constraints depends on the coupling realizations as well. 
For instance, if for a given realization, ${\sf C_1}, {\sf C_2}$ and ${\sf C_3}$ are satisfied,
trivially from Eq. \eqref{Energy_dif_SM} (i.e., $\Delta E_{a,e} = \Delta E_{a,b} + \Delta E_{a,d}$),
${\sf C_4}$ is automatically satisfied.
Therefore, for such cases, ${\sf C_4}$ is a linear combination of  ${\sf C_1}$ and ${\sf C_3}$,
and one may define the subset ${\overline{\sf Sat}}^{({\sf I})}_{| {\sf{s}} \rangle}$ 
for which ${\sf C_1}, {\sf C_2},{\sf C_3} \in {\overline{\sf Sat}}^{({\sf I})}_{| {\sf{s}} \rangle}$,
but ${\sf C_4} \notin {\overline{\sf Sat}}^{({\sf I})}_{| {\sf{s}} \rangle}$.
On the other hand, there exist some realizations for which $\Delta E_{a,b} = -  \Delta E_{a,d} \neq 0$,
but $\Delta E_{a,e} = 0$. Meaning, ${\sf C_4}$ is satisfied, however, ${\sf C_1}$ and ${\sf C_3}$ are not. 

The geometrically disjoint constraints may also give rise to different subsets. 
For instance, from Fig.~\ref{fig:GRC_1}, one can trivially show that
the pairs ${\sf C_1}, {\sf C_2}$ and ${\sf C_1}, {\sf C_3}$ are each geometrically disjoint, 
however, ${\sf C_2}$ and ${\sf C_3}$ are not. Therefore, we could define two different 
subsets ${\overline{\sf Sat^{g}}}^{({\sf I})}_{| {\sf{s}} \rangle}$
and ${\overline{\sf Sat^{g}}}^{({\sf II})}_{| {\sf{s}} \rangle}$ so that
${\sf C_1}, {\sf C_2} \in {\overline{\sf Sat^{g}}}^{({\sf I})}_{| {\sf{s}} \rangle}$
and ${\sf C_1}, {\sf C_3} \in {\overline{\sf Sat^{g}}}^{({\sf II})}_{| {\sf{s}} \rangle}$.

These examples further illustrate the difference between ${\cal M}$ and $M(\{ {\cal J}^m_{\alpha} \})$ 
in the main text, where ${\cal M}$ is associated with the maximum number of linearly independent 
satisfied constraints, i.e., the cardinality of ${\overline{\sf Sat}}_{| {\sf{s}} \rangle}$, 
while $M(\{ {\cal J}^m_{\alpha} \})$ denotes the number of constraints satisfied for 
a particular realization of coupling constants.
Trivially, $M(\{ {\cal J}^m_{\alpha} \}) \leq {\cal M}$.

\newpage
\onecolumngrid
$~~~$

$~~~$

\twocolumngrid

\subsection{The meaning of  Equation \eqref{binary} of the main text}
\label{the_meaning_of_it_all}

\begin{figure}
  \includegraphics[width= \columnwidth]{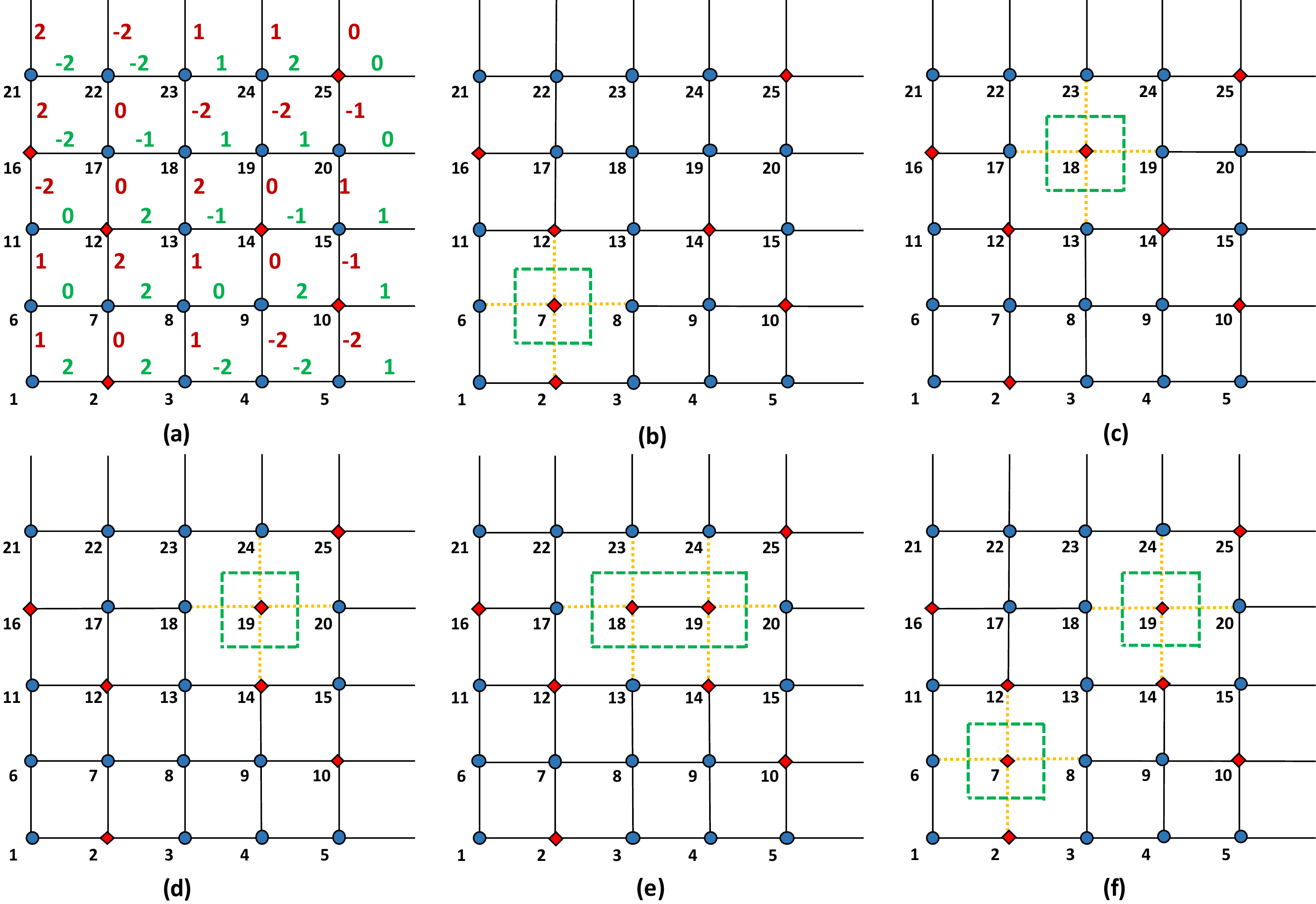}
  \caption{ 
  Panel (a) represents a random spin configuration with some given coupling constants.
  Blue solid circles and red diamonds denote spin up and down, respectively.
  The numbers in green (brown) color provide the values of horizontal (vertical) coupling constants.
  Flipping one or more spins at different sites of panel (a) would result in new spin configurations such as in panels (b) through (f).
  The dashed yellow dotted lines represent the links that contribute to the energy difference.
  The green dashed lines crossing such links correspond to a domain wall.
  Please note that the values associate with different links in each panel is the same as in panel (a).}
  \label{fig:Ilum_6}
\end{figure}

In Eq.~\eqref{binary} of the main text, we mentioned that the set
$\{ |\bar{n}_{1} \bar{n}_{2} \cdots \bar{n}_{M} \rangle \}$ includes all of the spin
configurations degenerate with $|{\sf s} \rangle$.  We also pointed out that it may
contain additional states not degenerate with $|{\sf s} \rangle$.  The latter point
is usually associated with the domain walls that are not geometrically disjoint (see
section \ref{GRC}).  To accentuate this consider, e.g., a $5\times 5$ lattice with a
given random spin configuration and coupling constants (see panel (a) of Fig.~
\ref{fig:Ilum_6}), in which $U_{{\sf ba}}, U_{{\sf ca}}$, and $U_{{\sf da}}$ are spin
flip operators leading, respectively, to zero energy domain walls around the sites
$7,18$ and $19$ (corresponding to panels (b),(c) and (d)).

From Fig.~\ref{fig:Ilum_6}, the domain walls in panel (c) and (d) are not geometrically disjoint,
where $U_{{\sf ca}}$ and  $U_{{\sf da}}$ act on the {\it nearest neighbor} sites $18$ and $19$
such that the sign of the link connecting them, is altered by both operators. 
In such a case, even though the two states $U_{\sf ca}| {\sf a} \rangle \equiv |{\sf c}\rangle $
and $U_{\sf da}| {\sf a} \rangle \equiv |{\sf d}\rangle $ are degenerate with $| {\sf a} \rangle$,
the state $U_{\sf da}U_{\sf ca}  | {\sf a} \rangle \equiv |{\sf e} \rangle$ (i.e., from panel (e), $U_{\sf ea} = U_{\sf da}U_{\sf ca}$)
is not degenerate with $| {\sf a} \rangle$.
One should note that in general this might not be true. That is, for some coupling realizations
the state $| {\sf e} \rangle$ can be degenerate with $| {\sf a} \rangle$.

By contrast, the two spin flip operators $U_{{\sf ba}}$ and $U_{{\sf da}}$
associated with the geometrically disjoint domain walls in panel (b) and (d), respectively, 
do not alter the signs of any common links. Therefore,
the state $U_{\sf da}U_{\sf ba}  | {\sf a} \rangle \equiv |{\sf f} \rangle$ (i.e., from panel (f), $U_{\sf fa} = U_{\sf da}U_{\sf ba}$)
is degenerate with $| {\sf a} \rangle$.

\newpage
\onecolumngrid
$~~~$
\twocolumngrid

\subsection{The ground state entropy is bounded by the entropy of a random energy level}
\label{gsbound}

In deriving the bound of Eq. (\ref{final!}) of the main text, we assumed that no information other than the probability 
distribution $P(\{ {\cal J}^{m}_\alpha \})$ is provided. The configuration $| {\sf{s}} \rangle$ that we considered in the main text
was an arbitrary random state. We next consider a more sophisticated problem. Suppose that the coupling constants are drawn from a binomial distribution and that once chosen
a {\it ground state configuration} ${| {\sf{s}} \rangle}$ is given (i.e., the values of the spins $s_{\sf{x}}$ at all sites ${\sf x}$ in this ground state are provided).
We then calculate the average of Eq. (\ref{nm}) of the main text with the condition that the (otherwise random binomial) coupling constants 
admit the particular configuration ${| {\sf{s}} \rangle}$ as {\it a ground state}.  When applicable, the fact that $| {\sf{s}} \rangle$ is a ground state may generally yield nontrivial constraints on the coupling constants $\{J_{\alpha}^{(k)}\}$ (recall that ${\cal J}^{m}_\alpha \equiv \frac{1}{\sqrt{m}} \sum^{m}_{k=1} J^{(k)}_\alpha$). 
In such a situation, given the configuration $| {\sf{s}} \rangle$, 
we may not simply use the initial binomial distribution for the coupling constants.

We now trivially demonstrate that if the energy density associated with the high temperature limit is unique then 
Eq. (\ref{final!}) of the main text will constitute an upper bound on the average ground state entropy density even if such information was provided for 
each realization of $\{J_{\alpha}^{(k)}\}$. This assertion follows as  the entropy $S_{\ell}(\{J_{\alpha}^{(k)}\})$ associated with any energy $E=E_{\ell}$ is typically larger than the ground state entropy, 
\begin{eqnarray}
\label{sgss}
S_{0} \le S_{\ell}.
\end{eqnarray}
The proof of Eq. (\ref{sgss}) is rather elementary and relies on a trivial symmetry of the spectrum. 
Let us denote the two sublattices forming the large bipartite lattice by $A$ and $B$.
If we flip all spins in sublattice $A$ (i.e., $s_{\sf{x}\in A} \to - s_{\sf{x} \in A}$) and do not
alter those in sublattice $B$ ($s_{\sf{y}\in B} \to s_{\sf{y}\in B}$), then all nearest-neighbor links
(i.e., the products $s_{\sf{x}} s_{\sf{y}}$ for nearest neighbor sites ${\sf{x}}$ and ${\sf{y}}$)
on the original lattice change 
their sign, $z_{\alpha} \to -z_{\alpha}$.
This single sublattice spin inversion constitutes a one-to-one mapping of the Ising spin states, 
that changes the sign of the total energy ($E \to -E$). We may thus conclude that
as a function of the energy $E$,  the entropy density ${\cal S} = S(\{J_{\alpha}^{(k)}\})/N$  for a system with fixed couplings $\{J_{\alpha}^{(k)}\}$
 satisfies the simple relation ${\cal S}(E_{\ell}) = {\cal S}(-E_{\ell})$ where $E_{\ell}$ is the energy of the 
$\ell$-th level. It follows that the energy $E=0$ is an extremum of the entropy
density ${\cal S}(E) \equiv {\cal S}(E_{\ell})$. Consequently, for any fixed couplings $\{J_{\alpha}^{(k)}\}$, 
\begin{eqnarray}
\label{1ts}
\frac{1}{T} = N \frac{\partial {\cal S}}{\partial E}  \ge 0.
\end{eqnarray}
(The factor of $N$ appears in the above equation since ${\cal S}$ is the entropy density). 
Thus, $E \le 0$ for any positive temperature $T$. In what follows we discuss what occurs if there is a {\it unique} high temperature limit for each set of coupling 
constants. In such a case, the entropy density ${\cal S}(E)$ (averaged over all realization of the coupling constants) is maximal at $E=0$. The 
semi-positive definite nature of the derivative in Eq. (\ref{1ts}) implies (as in all common systems satisfying the third law of 
thermodynamics) that the entropy is lowest at $T=0$. Since the state $| {\sf{s}} \rangle$ for which we  performed the analysis was
arbitrary (and corresponds to an energy $E_{| {\sf{s}} \rangle}$ for which the entropy density is greater than or equal to that of the ground state), 
we see that Eq. (\ref{1ts}) must hold even if information is provided as to the explicit ground state configuration $| {\sf s} \rangle$ for each particular 
realization of the couplings $\{J_{\alpha}^{(k)}\}$. We thus observe that even if given such additional information, 
the ground state entropy density must satisfy the bound of Eq. (\ref{final!}) of the main text. 

\subsection{Asymptotic Scaling of the Entropy Density}
\label{asymptotics}

We now motivate a scaling that suggests that the rigorous bound of Eq. (\ref{final!}) of the main text leads to Eq. (\ref{final-conjecture}) 
as an approximate asymptotic relation for large $N$ and $m$. 
In Section \ref{GRC} of this supplemental material, we defined
the subset ${\overline{\sf Sat^{g}}}_{| {\sf{s}} \rangle} \subset
{\overline{\sf Sat}}_{| {\sf{s}} \rangle}$ composed of geometrically disjoint constraints.
 If there are $n_{g}$ such constraints (or associated zero energy domain walls when these constraints are satisfied)
 then the degeneracy will be trivially bounded from below by $2^{n_{g}}$. 
 This bound is established by noting that, since no spin is common to two domain walls, all of the spins in 
 each of these $n_{g}$ domain walls may 
be flipped independently of all others. When applied to domain walls in 
 ${\overline{\sf Sat^{g}}}_{| {\sf{s}} \rangle}$ then, in the notation of Eq. (\ref{binary}) of the main text, each binary string
 of length $n_{g}$ will correspond to
 a different configuration that is degenerate with the reference state $|{\sf s} \rangle$. 
 This is to be contrasted with the set of zero energy domain walls ${\overline{\sf Sat}}_{| {\sf{s}} \rangle}$
 for which various binary strings of the form of Eq. (\ref{binary}) may correspond to
 states that are {\em not} degenerate with $| {\sf s} \rangle$. As $m$ grows, by 
 Eq. (\ref{m-scale}) of the main text, both the number of satisfied constraints 
and the number of independent zero energy domain walls may diminish as $1/\sqrt{m}$. 
When fewer walls appear in ${\overline{\sf Sat}}_{| {\sf{s}} \rangle}$, it may become increasingly rare for
different walls in this subset to share the same lattice sites. If this occurs then, for large $m$, 
we will have the asymptotic relation $ {\overline{\sf Sat^{g}}}_{| {\sf{s}} \rangle} \sim
 {\overline{\sf Sat}}_{| {\sf{s}} \rangle}$. In such a case, in the large $N$ limit, 
 ${\cal S} \sim n_g/N \ln 2$. The number $n_{g}$ and the probability of these zero energy domain walls 
 decay, for $m \gg 1$, as $1/\sqrt{m}$ (or $1/\sqrt{m'}$ for $m' \gg 1$).
Similarly, if a finite fraction of the $M$ domain walls
 in ${\overline{\sf Sat}}_{| {\sf{s}} \rangle}$ does not remain geometrically disjoint such that, asymptotically, 
 one may only generate $q^{M}$ 
 (with $q<2$) degenerate states (Eq. (\ref{binary})) given $M$ independent domain walls, then 
 ${\cal S} \sim \frac{M}{N} \ln q$. Either way, we anticipate that, in the thermodynamic limit,
Eq. (\ref{final-conjecture}) of the main text will hold.

\subsection{One-dimensional Binomial Spin Glass}
\label{1dEA+}

Let us start with the simplest one-dimensional binomial spin glass system (which by a simple change of variables
($s_{\sf x} \to s'_{\sf x} \equiv s_{\sf x} \prod_{\sf{u} < {\sf x}} {\sf sign}({\cal J}^{m}_{{\sf u,u+1}})$)
may be transformed onto a random Ising ferromagnet with couplings $|{\cal J}_{\sf{x',x'+1}}^{m}|$).
Here, the ground state energy $E_{0}= - \sum_{x} |{\cal J}^{m}_{\sf{x,x+1}}|$. 
In an open  chain of $N$ sites, the lowest excitation 
consists of identifying the weakest link, $|{\cal J}^{m}_{\sf{x',x'+1}}| \equiv \min_{\sf{x}} \{ |{\cal J}^{m}_{\sf{x,x+1}}|\}$ 
and flipping all spins $s_{\sf{x}} \to - s_{\sf{x}}$ for which ${\sf{x}}>{\sf{x'}}$ (or consistently doing the same thing 
and only flipping all spins to the left of ${\sf{x'}}$); this generates a state that has an energy $E_0 +  \Delta E_{\min}$
with $\Delta E_{\min} = 2|{\cal J}^{m}_{\sf{x',x'+1}}|$. (On a periodic chain, we may similarly identify 
the two weakest links and flip all spins lying between those two links leading to an energy cost $\Delta E_{\min}$ 
that is twice the sum of the moduli of these two weakest links.) Calculations of the density 
of states and all ensuing thermodynamic properties are trivial \cite{trivial}.
For instance, the disorder averaged entropy  in the low 
temperature, $T \ll 1$, limit of the binary model is $[S_{m=1}(T)] \sim k_{B} (\ln 2+ (N-1) (1+ 2 \beta) e^{-2 \beta})$, 
with $\beta= 1/(k_{B} T)$. 
The exponential suppression becomes $e^{-2 \beta/\sqrt{m}}$ and $e^{-4 \beta/\sqrt{m}}$ for odd and even $m$, 
respectively. Thus the excitation gap scales as $m^{-1/2}$ ({\it yet differently for odd and even} $m$).
By contrast, the low-$T$ entropy of the continuum  
model is $[S_{m \to \infty}(T)] \sim k_{B} (\ln 2 +   \frac{N-1}{\sqrt{2\pi}} (k_{B}T - \frac{(k_{B} T)^{3}}{8}))$, 
indicating the vanishing of the spectral gap in the thermodynamic limit. In that limit,
these lowest excitations differ, 
relative to the ground state, by 
an extensive number of flipped spins.

\subsection{Distribution of excitations}
\label{distr1}

Given any {\it ground state} configuration on a hypercubic lattice in $d$ dimensions, one may compute the probability distribution for excitations 
of energy $\Delta E_{\sf x} = |\Delta E_{\sf x}| = 2  \sum_{\sf y} n_{\sf{xy}} {\cal J}^{m}_{\alpha}$
generated by flipping a single spin ${\sf x}$. Here, the sum is over all sites ${\sf y}$ that are nearest 
neighbor of site ${\sf x}$ and $n_{\sf{x y}} = - {\sf sign}(s_{\sf x} s_{\sf y}) = \pm 1$. Given the probability 
distribution for the links $\{{\cal J}^{m}_{\alpha}\}$, one may compute the probability distribution 
associated with a finite sum of these links
$2  \sum_{\sf y} n_{\sf{xy}} {\cal J}^{m}_{\alpha}$ in the ground state. The latter sum is that over a finite 
number of links (with bounded mean and variance)
and thus for any $\epsilon>0$ (no matter how small), the probability that $|\Delta E_{\sf x}| < \epsilon $ is strictly smaller than unity.
In order for the system to have a spectral gap that is greater than $\epsilon$, it must be that {\it for each of the $N$ lattice sites} ${\sf x}$,
the energy penalty $|\Delta E_{\sf x}| >\epsilon$. Given that the condition $|\Delta E_{\sf x}| >\epsilon$ must, in the thermodynamic limit, 
be satisfied an infinite number of times, while for any single ${\sf x}$ the probability that this condition is satisfied is strictly smaller than one,
it is essentially impossible to have a gap larger than any arbitrary positive number $\epsilon$. From this, it 
follows that the gapless local excitations must be appear.
If the local energy penalties in the ground state were independent of one another then the 
probability that all local flips result in an energy penalty larger than $\epsilon$ would the product of the 
probabilities of having $|\Delta E_{\sf x}| >\epsilon$ for all sites ${\sf x}$. Although the local flip are not independent of one another
(since they all relate to flips relative to the same special state- the ground state), it seems highly unlikely 
$|\Delta E_{\sf x}| >\epsilon$ for all ${\sf x}$ when the probability of having a local energy penalty larger 
than $\epsilon$ for any single ${\sf x}$ is strictly smaller than one.

\begin{figure}
  \includegraphics[width=\linewidth]{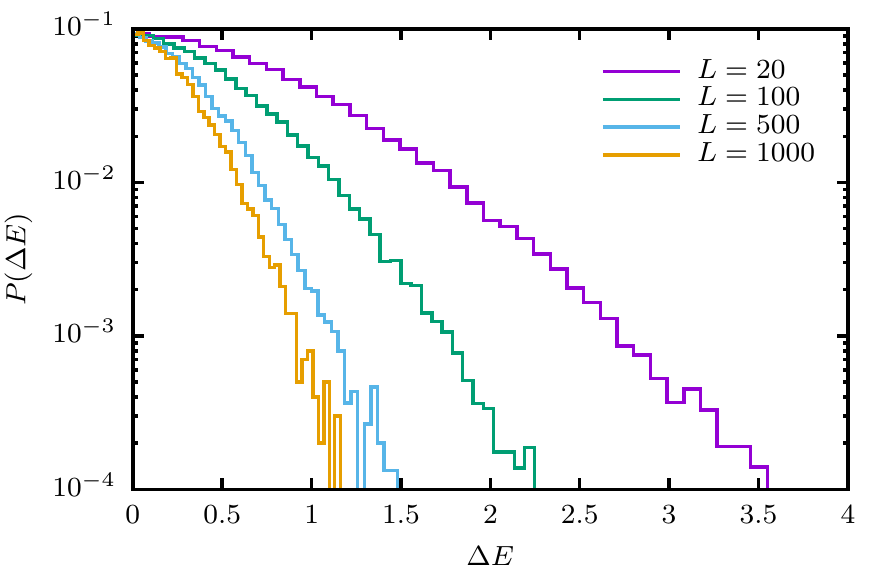}
  \caption{ Distribution of defect energies $|\Delta E|$ for the 2D Gaussian system
    and a number of different system sizes. The curves collapse if rescaled by
    $L^{-\theta}$ with the value $\theta \approx -0.28$ of the stiffness exponent.
  }
  \label{fig:distribution}
\end{figure}

We now explicitly discuss a measure that, in general dimensions, may provide physical insight -- the {\it distribution} of such individual
defect energies (i.e., the distribution of domain wall energies in our binomial Ising spin
system). In Fig.~\ref{fig:distribution}, we plot this distribution in the continuous $m = \infty$ Gaussian limit. 
If $f(\epsilon,\tilde{l})$ denotes
the cumulative probability that the energy penalty of a domain wall (of size
$\tilde{l}$) is smaller than $\epsilon$, then the probability that amongst
${\cal{N}}_{\tilde{l}}$ {\it independent} domain walls, no singe domain wall entails
an energy cost lower than $\epsilon$ will be bounded from above by
$e^{-f(\epsilon,\tilde{l}) {\cal{N}}_{\tilde{l}}}$ as we briefly elaborate on now.  
Since, by definition,  $f(\epsilon,\tilde{l})$ is the cumulative probability that the energy cost of a random wall 
of size $\tilde{l}$ is smaller than $\epsilon$ (i.e., $Prob.(|\Delta E| \le \epsilon) =  f(\epsilon, \overline{l})$), the probability
that amongst ${\cal{N}}_{\tilde{l}}$ {\it independent} domain walls, we explicitly have that the probability that no single domain wall 
has an energy cost larger than $\epsilon$ is, trivially, $[Prob. (|\Delta E|> \epsilon)]^{{\cal{N}}_{\tilde{l}}} = 
(1-f(\epsilon, \tilde{l}))^{{\cal{N}}_{\tilde{l}}} \le e^{-{\cal{N}}_{\tilde{l}} f(\epsilon, \tilde{l})}$ (where we invoked $e^{-f} \ge (1-f)$ for all $f \ge 0$). 
For small $f \to 0^{+}$ (associated with $\epsilon \to 0^{+}$ in $d\ge 3$), this general inequality is replaced by an equality
(i.e., $[Prob. (|\Delta E|> \epsilon)]^{{\cal{N}}_{\tilde{l}}} = e^{-{\cal{N}}_{\tilde{l}} f(\epsilon, \tilde{l})}$).

 Thus, if the area
($d=2$) or volume $(d=3)$ of the entire lattice is $||\Lambda||$, then whenever the
sum
\begin{eqnarray}
\label{diverge}
\lim_{\epsilon \to 0^{+}} \lim_{\tilde{l}_{0} \to \infty} \lim_{N \to \infty} \sum_{||\Lambda||^{1/d}- \tilde{l}_{0} \ge  \tilde{l} 
\ge \tilde{l}_{0}} f(\epsilon,\tilde{l}) {\cal{N}}_{\tilde{l}} = \infty
\end{eqnarray}
then gapless (or degenerate) states of diverging $\tilde{l}$ may appear. This is so because flipping all of the spins links one ground state 
to its conjugate. The inequality $||\Lambda||^{1/d}- \tilde{l}_{0} \ge  \tilde{l} 
\ge \tilde{l}_{0}$ in Eq. (\ref{diverge}) means that the an extensive number of spin flips is needed to 
connect a given spin configuration to either of the two members of the degenerate ground state pair. 

Since $\theta_{d=2} <0$ then (as is further underscored in the full
distribution of Fig.~\ref{fig:distribution}), in two dimensions nearly all large
domain walls entail a {\it vanishing} energy penalty. In $d=2$,
$\lim_{\epsilon \to 0^{+}} \lim_{{\tilde{l}} \to \infty} f(\epsilon, \tilde{l}) =1$
and the probability of obtaining, in the thermodynamic limit, degenerate
states that differ by an extensive number of flipped spins is unity. 
The existence of gapless states in $d=2$ is hardly surprising; such
gapless states may be trivially constructed by the insertion of random domain walls
of divergent size into a ground state. Indeed, in $d=2$ (where the typical energy
cost ${\cal{O}}(\tilde{l}^{\theta})$ vanishes as $\tilde{l} \to \infty$), knowledge
of the detailed distribution of the energy cost as a function of the domain wall size
$\tilde{l}$ is unnecessary for establishing gapless states. However, in
$d \ge 3$ (where $\theta_{d} >0$), the lowest energy states are related
to {\it the asymptotic low energy limit} of the domain wall energy distribution (a
distribution that, in these higher dimensions, is associated with a divergent average
energy ${\cal{O}}(\tilde{l}^{\theta_{d}})$ when $\tilde{l} \to \infty$). A gap (for
states that differ from one another by an extensive number of flipped spins) is potentially possible if the sum of Eq. (\ref{diverge})
vanishes. Thus, we stress that in $d \ge 3$, knowledge of the cumulative probability
distribution $f(\epsilon, \tilde{l})$ can be of paramount importance. We reserve the
analysis of the $d=3$ domain wall energy distribution for future work.


\begin{thebibliography}{28}
\bibitem{def:incongruent}  
Given any Ising spin configuration one may inspect the sign of each of the links
$z_{\alpha}$ on the lattice.  If there is
an extensive (volume proportional) number of links $z_{\alpha}$ that are of different signs in two different Ising
spin configurations  $|{\sf{s}}\rangle$ and $|{\sf{s'}}\rangle$ then the two states 
are said to be ``incongruent'' relative to one another \cite{Stein-Complexity}.
\end{thebibliography}
\begin{thebibliography}{28}
\bibitem{trivial} 
In order to attain any state, a number 
of links $\{{\cal J}^{m}_{\alpha}\}$ must be flipped relative to a ground state. 
The probability for obtaining a particular energy amounts to a convolution on the 
probability distribution of the links. The latter convolution readily becomes a trivial product after Fourier transformation.
\end{thebibliography}
\end{document}